\documentclass[aps,pra,reprint,groupedaddress]{revtex4-2}

\usepackage{amsmath}
\usepackage{dcolumn}
\usepackage{graphicx}
\usepackage{xcolor}
\usepackage{siunitx}
\usepackage{hyperref}

\newcommand{\Hop}{\hat{H}}
\newcommand{\bop}{\hat{b}}
\newcommand{\bopd}{\hat{b}^\dag}
\newcommand{\nop}{\hat{n}}

\begin{document}

\title{Quantum phases of bosonic chiral molecules in helicity lattices}

\author{Felipe Isaule}
\email[]{felipe.isaulerodriguez@glasgow.ac.uk}
\author{Robert Bennett}
\author{J\"org B. G\"otte}
\affiliation{School of Physics and Astronomy, University of Glasgow, Glasgow G12 8QQ,
United Kingdom}

\date{\today}

\begin{abstract}
We reveal the existence of polarizing quantum phases for the enantiomers of cold, interacting chiral molecules in an optical helicity lattice by means of an extended Bose-Hubbard model.
These recently proposed lattices have
sites with alternating helicity which exert a discriminatory force on chiral molecules with different handedness. 
In our study of the phase diagram we find that a strong dipolar repulsion between molecules results in the separation of left and right enantiomers.
\end{abstract}

\maketitle

\section{Introduction}

Chirality is a near-universal phenomenon in the natural world, with the most basic building blocks of life consisting of chiral molecules~\cite{blackmond2011}. It is a geometrical property, with chiral objects being those that cannot be brought into coincidence with their mirror image by rotations and translations, much like left and right hands. The ability to identify and separate molecules of different chirality (known as enantiomers) is of crucial importance in pharmaceuticals. A variety of ways of achieving this have been proposed \cite{patterson2014}, most of which rely on circular dichroism spectroscopy \cite{bowering2001}. However, such a method comes with some distinct disadvantages, the most serious of which is that the signals are so weak that the sample must be highly concentrated and in the liquid phase \cite{blackmond2011}. Partly due to this, there has been a continued interest in alternative ways of identifying and separating chiral molecules, including Coulomb explosion imaging \cite{pitzer2013}, three-wave mixing \cite{patterson2013}, mechanical optical forces \cite{canaguier-durand_mechanical_2013,cameron_discriminatory_2014} and even phase transitions in quantum fluids of light \cite{bennett2020a}.  

\begin{figure}[t]
\centering
\includegraphics[scale=0.4]{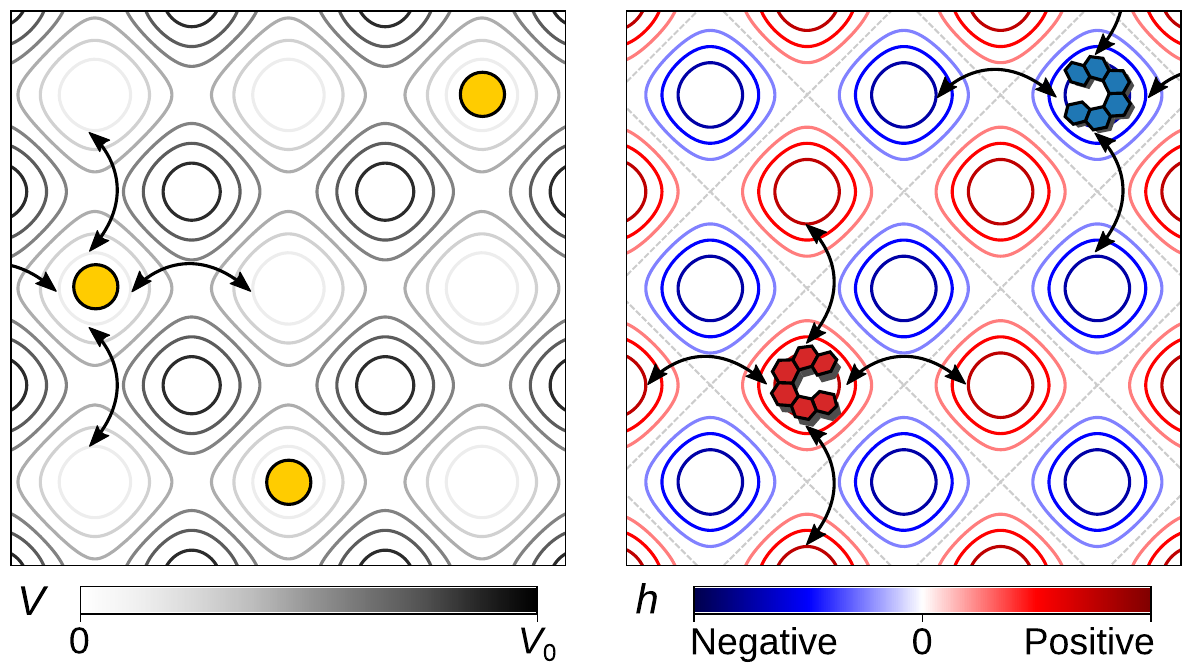}
\caption{(Left): Illustration of a conventional optical lattice. Oscillating electric fields induce a periodic potential for atoms (represented as yellow circles). We show an homogeneous lattice potential $V=V_0(\sin^2(kx)+\sin^2(ky))$, where $k$
depends on the laser’s wavelength~\cite{bloch_many-body_2008}.
(Right): Illustration of a square helicity lattice. Red and blue regions correspond to sites with opposite helicity densities $h$~\cite{van_kruining_superpositions_2018} which induce a
discriminatory force on enantiomers~\cite{canaguier-durand_mechanical_2013}.}
\label{sec:intro;fig:lattice}
\end{figure}

Here we take an entirely different approach, exploiting recently-proposed \emph{helicity lattices}~\cite{van_kruining_superpositions_2018}. These are optical lattices that are engineered in such a way that they have perfectly homogeneous mean squared values of the electric field, but spatially varying helicity [see Fig.~\ref{sec:intro;fig:lattice}]. This means that chiral molecules within the lattice will have dynamics induced by their chirality to leading order, perhaps enabling new methods of separation and characterization of mixtures. 

Recent progress on laser cooling~\cite{mccarron_laser_2018,tarbutt_laser_2018,isaev_direct_2020,vilas_magneto-optical_2022} has brought molecules to the ultracold regime~\cite{anderegg_laser_2018}. The cooling of a growing selection of polyatomic molecules has started to be achieved~\cite{klos_prospects_2020}, including the recent cooling of CaOCH$_3$ molecules~\cite{mitra_direct_2020}. In particular, roadmaps for cooling and trapping chiral molecules have already been proposed~\cite{isaev_polyatomic_2016,augenbraun_molecular_2020}. This opens the possibility of trapping cold chiral molecules in helicity lattices in the next few years. At such low temperatures, helicity lattices would show novel quantum phases, in analogy to cold atomic gases trapped in conventional optical lattices~\cite{gross_quantum_2017}. In addition, Rydberg atoms with induced chirality have recently been proposed~\cite{buhmann_quantum_2021}, potentially offering a more controllable scenario for simulating ultracold chiral particles.

In this work we consider an idealized system of interacting chiral molecules immersed in a two-dimensional helicity lattice. We model the lattice with a modified extended Bose-Hubbard model~\cite{dutta_non-standard_2015} where the chiral molecules are represented as structureless bosonic particles. We tune parameters such as the dipole-dipole coupling strength between molecules and tunneling rate to identify a number of new phases which are unique to helicity lattices. This work forms a proof-of-principle which should be achievable in near-future molecular cooling experiments.

\section{Model}
\label{sec:model}

We consider a system of left and right enantiomers of the same chiral molecule immersed in a square helicity lattice as depicted in the right panel of Fig.~\ref{sec:intro;fig:lattice}. Neighboring sites have opposite helicity densities $h$~\cite{van_kruining_superpositions_2018}, simulating conditions analogous to those of a helicity lattice created by the superposition of four coherent light waves~\cite{van_kruining_superpositions_2018}.

The chirality of a particular enantiomer determines whether it is attracted or repelled from a given site~\cite{canaguier-durand_mechanical_2013}. Indeed, because the chiral forces acting on the enantiomers are proportional to $\nabla h$~\cite{cameron_discriminatory_2014,cameron_diffraction_2014}, the enantiomers are immersed in a periodic potential with wells at the sites with favorably helicity. 
For each enantiomer species, the corresponding potential depth is given by~\cite{cameron_discriminatory_2014, cameron_chirality_2017}
\begin{equation}
    V_0 \approx \frac{|G'| I}{\epsilon_0 c^2}\,,   
\end{equation}
where $G'$ is the relevant part of the electric dipole-magnetic dipole optical activity tensor~\cite{barron2009molecular} and $I$ is the laser power. $V_0$ is to be compared with the recoil energy $E_R=\hbar^2 k^2/2m$, where $m$ is the mass of the molecules and $k=2\pi/\lambda$ depends on the laser's wavelength $\lambda$. In this work we consider tight lattices with $E_R\ll V_0$ at low temperatures $k_BT\ll V_0$, which we can model with a Hubbard-based Hamiltonian. In addition, we consider ultracold regimes where the molecules are dominated by their ground rovibrational state~\cite{hamamda_ro-vibrational_2015,mitra_direct_2020}.

The previous assumptions set constraints on our proposed experimental setup. To trap hexahelicene ($|G'|\approx 0.33\times10^{-34}\text{m}\, \text{s}^3\,\text{A}^2\,\text{kg}^{-1}$) within a helicity lattice in the \textmu K regime we require a laser power of $I=10^9\mathrm{W/cm}^2$~\cite{cameron_chirality_2017}. In a lattice with $\lambda = 1\mu$m, the recoil energy takes a value of $E_R = 4.0\times 10^{-31}$J, while the potential depth $V_0=4.2\times10^{-28}\mathrm{J}\approx 10^3 E_R$. Other choices can be used to realize shallower potentials.

By neglecting rovibrational excitations, we can model the enantiomers as point-like bosonic particles, and thus, we can describe the system with a modified extended Bose-Hubbard model~\cite{dutta_non-standard_2015}
\begin{equation}
    \Hop=\Hop_{\text{(hop)}}+\Hop_{\text{(int)}}\,.
    \label{sec:model;eq:H}
\end{equation}
The hopping of the enantiomers is described by
\begin{equation}
    \Hop_{\text{(hop)}} = -\frac{t}{2}\sum_{\langle\langle i,j \rangle\rangle_\chi}\left(\bopd_{\chi,i}\bop_{\chi,j}+\bopd_{\chi,i}\bop_{\chi,j}\right)\,,
    \label{sec:model;eq:Hhop}
\end{equation}
where $t$ encodes the probability of tunneling to a neighboring site with favorable helicity, $\bopd_{\chi,i}$ ($\bop_{\chi,i}$) creates (annihilates) an enantiomer of handedness $\chi=L,R$ at site $i$, and $\langle\langle i,j \rangle\rangle_\chi$ denotes the next-to-nearest neighbor sites, that is, the nearest sites with the same helicity. We stress
that we allow the lattice to have left or right enantiomers
at alternating sites. The tunneling $t$ contains the effect of both the mass of the particles and of the depth of the lattice potential~\cite{dutta_non-standard_2015}. Note that because we consider enantiomers of the same chiral molecule, both enantiomers have equal masses and therefore $t=t_L=t_R$.

Chiral molecules interact through short-range van der Waals and long-range dipole-dipole interactions~\cite{craig_new_1999,salam_effect_2006}. By considering that all the dipoles are polarized orthogonal to the lattice-plane, the dipolar interactions are repulsive and take the simple form $V^{\text{(dip)}}_{\chi\chi'} = \tilde{V}_{\chi\chi'}/r^3$ for $\chi,\chi'=L,R$, where $r$ is the distance between dipoles and $\tilde{V}>0$ characterizes the strength of the dipolar repulsion. In the following, we assume equal dipolar repulsion between enantiomers of the same handedness $\tilde{V}=\tilde{V}_{LL}=\tilde{V}_{RR}$.

In this work, we consider dipolar interactions up to the next-to-nearest neighbors in order to capture the long-range interactions between enantiomers with opposite handedness. Therefore, the interacting part of the Hamiltonian reads~\cite{dutta_non-standard_2015}
\begin{align}
    \Hop_{\text{(int)}}=&\frac{U}{2}\sum_i \nop_{\chi,i}(\nop_{\chi,i}-1)+\frac{V_{LR}}{2}\sum_{\langle i,j \rangle_{LR}}\nop_{\chi,i}\nop_{\chi',j}\nonumber\\&+\frac{V}{2^{5/2}}\sum_{\langle\langle i,j \rangle\rangle_\chi}\nop_{\chi,i}\nop_{\chi,j}\,,
    \label{sec:model;eq:Hint}
\end{align}
where $\nop_{\chi,i}=\bopd_{\chi,i}\bop_{\chi,i}$ is the number operator and $\langle i,j \rangle_{LR}$ denotes the nearest neighbor sites, that is, the nearest sites with opposite helicity. The first term on the right-hand side corresponds to the on-site contact repulsion, the second term to the long-range dipolar repulsion between left and right enantiomers, and the third term to the long-range dipolar repulsion between enantiomers of the same handedness where the $2^{3/2}$ factor comes from the distance between next-to-nearest neighbors. We note that similar Hamiltonians have been used to study two-component Bose-Hubbard models~\cite{altman_phase_2003,isacsson_superfluid-insulator_2005}. However, most efforts have focused instead on attractive inter-species interactions~\cite{kuklov_commensurate_2004,kuklov_superfluid-superfluid_2004,li_anisotropic_2013}.

To study the phase diagram of Hamiltonian~(\ref{sec:model;eq:H}) we employ a Gutzwiller ansatz~\cite{rokhsar_gutzwiller_1991,jaksch_cold_1998}
\begin{equation}
    |\Psi(t)\rangle=\prod_i\sum_{m=0}^{m_\text{max}}f_{\chi,m}^{(i)}(t)|m\rangle_{\chi,i}\,,
    \label{sec:model;eq:Ansatz}
\end{equation}
where $|m\rangle_{\chi,i}$ denotes the state with $m$ particles of enantiomer $\chi$ at site $i$ and  $m_\text{max}$ is the maximum occupancy allowed in the numerical calculations. 

In this work we study the ground-state phase diagram by obtaining the coefficients $f_{\chi,m}^{(i)}$ from the minimization of $\langle \Psi | \Hop-\sum_i\mu_\chi \nop_{\chi,i}|\Psi\rangle$, where the chemical potentials $\mu_\chi$ control the density of each enantiomer species. Here we work with equal chemical potentials for both enantiomers $\mu=\mu_L=\mu_R$. We identify the quantum phases by examining the values of the order parameter and average occupancy per site $\phi_{\chi,i}=\langle \Psi | \bop_{\chi,i} |\Psi\rangle$ and $n_{\chi,i}=\langle \Psi | \nop_{\chi,i} | \Psi\rangle$, respectively. We refer to Ref.~\cite{trefzger_ultracold_2011} for details about the Gutzwiller approach for similar models.

\section{Quantum phases}
\label{sec:phases}

Hubbard models show superfluid and insulator phases. An occupied site is superfluid if its order parameter 
 is finite ($\phi_{\chi,i}>0$), whereas a site is in an insulator state if its order parameter is zero ($\phi_{\chi,i}=0$). Moreover, while superfluid sites can have any positive and real occupation ($n_{\chi,i}>0$), insulator sites have an integer occupation $n_{\chi,i}=\nu$, where $\nu$ denotes a positive integer number. In addition, repulsive dipolar interactions induce \emph{checkerboard} phases with staggered occupations~\cite{goral_quantum_2002,kovrizhin_density_2005,menotti_metastable_2007,trefzger_ultracold_2008,iskin_route_2011,ohgoe_ground-state_2012}. In this case, the lattice can have unoccupied sites with $n_{\chi,i}=0$.

\begin{figure}[t]
\centering
\includegraphics[scale=0.19]{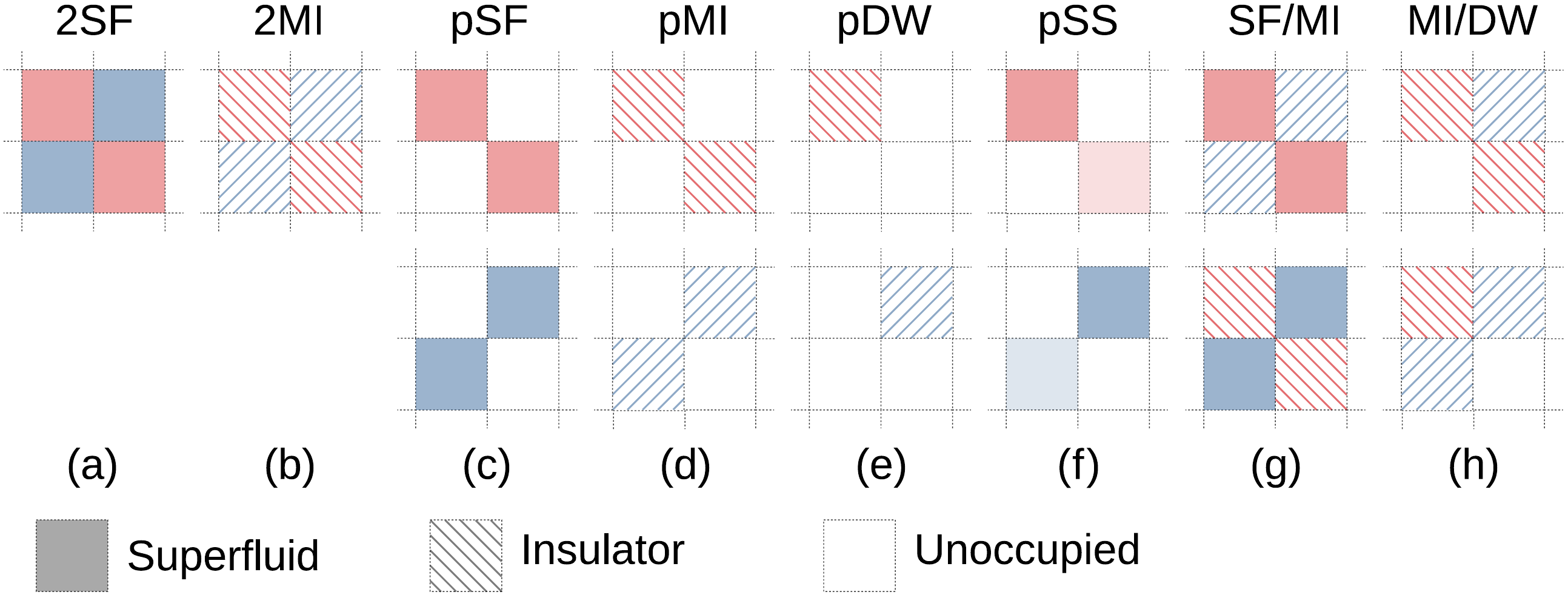}
\caption{Illustrations of a representative set of quantum phases. Red and blue squares correspond left and right sites, respectively. Filled squares correspond to superfluid sites, cross-hatched squares correspond to insulator sites, and blank squares correspond to unoccupied sites. The lattices in the second row correspond to equivalent degenerate phases by exchanging left with right sites.}
\label{sec:phases;fig:phases1}
\end{figure}

Because we consider up to next-to-nearest neighbor interactions, the system is sufficiently described 
by a 2$\times$2 lattice with periodic boundary conditions. Therefore, we work with two left ($L$) and two right ($R$) sites. We label the four sublattices as $L(R)_A$ and $L(R)_B$. We perform the minimization of the Gutzwiller ansatz (\ref{sec:model;eq:Ansatz}) over 4$\times\,(m_{\text{max}}+1)$  coefficients $f^{(i)}_{\chi,m}$, where $i$ corresponds to the four sublattices and $m=0,...,m_{\text{max}}$. Note that $\chi$ simply labels the handedness of the corresponding site. The specific quantum phase depends on the combination of  $\phi_{\chi,a}$ and $n_{\chi,a}$ in the four sublattices. We schematically illustrate the most prominent quantum phases in Fig.~\ref{sec:phases;fig:phases1}.

In the figure, phases (a) and (b) correspond to uniform configurations with equal order parameters and occupations in all sites. In this case, the left and right sublattices show either a Mott-insulator (MI) or a superfluid (SF) phase, and therefore we refer to these as 2MI and 2SF phases.

As mentioned, the presence of a dipolar repulsion between left and right enantiomers can produce a left/right polarization with $n_{L/R}>0$ and $n_{R/L}=0$. Depending on if the sublattice is a superfluid or an insulator, we refer to these phases as a polarized superfluid (pSF) or a polarized Mott-insulator (pMI) [(c) and (d)]. In addition, a dipolar repulsion between enantiomers of the same handedness can produce a further polarization within left/right sites [(e) and (f)]. We refer to these phases as polarized density-wave (pDW) and polarized supersolid (pSS) phases [(e) and (f)], in analogy to the crystalline DW and SS phases with staggered occupations of extended Bose-Hubbard models~\cite{kovrizhin_density_2005}.

The lattice can also show intermediate configurations, such as SF/MI [(g)] and MI/DW [(h)] phases, where one enantiomer species shows a MI phase and the other a SF or DW phase.
We present a complete list of the phases in Tables~\ref{sec:phases;table:phasesMI} and~\ref{sec:phases;table:phasesSF}, including generalizations of the phases illustrated in Fig.~\ref{sec:phases;fig:phases1} for arbitrary occupations. Note that $n$ and $\phi$ denote the total average occupation and order parameter, respectively.

\begin{table}[b]
\caption{Average occupations for the insulator phases, where $\nu$, $\nu'$, and $\nu''$ denote integer numbers. Note that in these phases the order parameters are zero.\label{sec:phases;table:phasesMI}
}
\begin{ruledtabular}
\begin{tabular}{lcccccc}
Phase & $n$ & $|n_L-n_R|$ & $|n_A-n_B|$\\
\colrule
2MI$_\nu$ & $\nu$ & $0$ & $0$\\
MI$_\nu$/MI$_{\nu'}$ & $(\nu+\nu')/2$ & $|\nu-\nu'|$ & $0$\\
pMI$_\nu$ & $\nu/2$ & $\nu$ & $0$\\
MI$_\nu$/DW$_{(\nu',\nu'')}$ & $\frac{\nu}{2}+\frac{\nu'+\nu''}{4}$ & $|\nu-\frac{\nu'+\nu''}{2}|$ & $|v'-v''|/2$\\
pDW$_{(\nu,\nu')}$ &  $(\nu+\nu')/4$ & $(\nu+\nu')/2$ & $|\nu-\nu'|/2$
\end{tabular}
\end{ruledtabular}
\end{table}

\begin{table}[b]
\caption{Average order parameters and occupations for the superfluid phases. Note that $n>0$ and $\phi>0$ in all phases.
\label{sec:phases;table:phasesSF}
}
\begin{ruledtabular}
\begin{tabular}{lcccccc}
Phase &  $|\phi_L-\phi_R|$ & $|\phi_A-\phi_R|$ &  $|n_L-n_R|$\\
\colrule
2SF &  $0$ & $0$ &  $0$\\
pSF &  $2\phi$ & $0$ &  $2n$\\
pSS &  $2\phi$ & $>0$ & $2n$\\
spSF &  $>0$ & $0$ & $>0$\\
MI$_{\nu}$/SF  & $2\phi$ & $0$ & $>0$\\
MI$_{\nu}$/SS  & $2\phi$ & $>0$ & $>0$
\end{tabular}
\end{ruledtabular}
\end{table}

\section{Phase diagram}
\label{sec:PD}

\begin{figure}[t]
\centering
\includegraphics[scale=0.8]{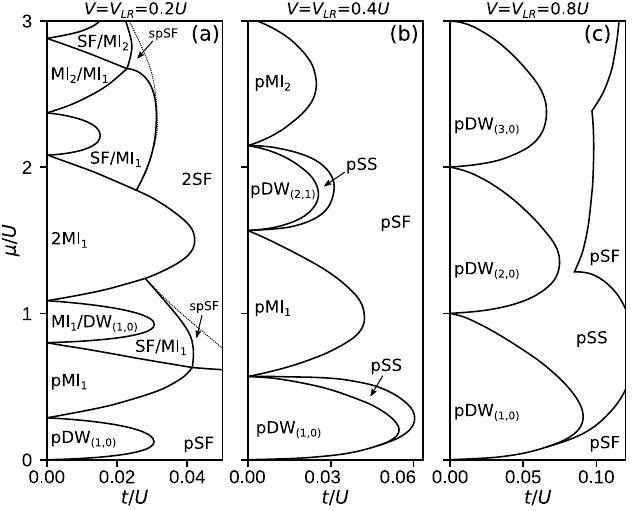}
\caption{Phase diagrams for $V=V_{LR}$ and $\mu_L=\mu_R$ as a function of the chemical potential $\mu/U$ and hopping $t/U$.}
\label{sec:PD;fig:PD1}
\end{figure}

\begin{figure}[t]
\centering
\includegraphics[scale=0.8]{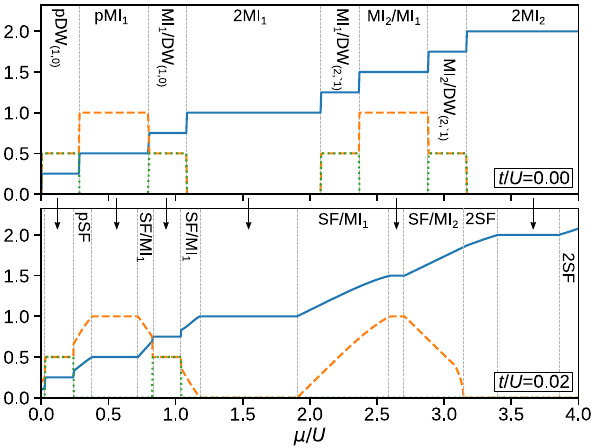}
\caption{Average occupations $n$ (blue solid lines), $|n_L-n_R| $ (dashed orange lines) and $|n_A-n_B| $ (dotted green lines) for $V=V_{LR}=0.2U$ as a function of $\mu/U$. We show results for $t=0$ (upper panel) and $t=0.02U$ (bottom panel).}
\label{sec:PD;fig:n_0o02}
\end{figure}

We first examine the phase diagram for different values of $V_{LR}=V$. We show a representative set in Fig.~\ref{sec:PD;fig:PD1}. The phase diagram is particularly rich for weak dipolar repulsion $V_{LR}<U/4$ and $V/2^{2/3}<U/4$ [see Fig.~\ref{sec:PD;fig:PD1}(a)]. Under these constraints, the on-site repulsion dominates over the long-range one, allowing both uniform and polarized configurations. 
Indeed, for small tunneling the system shows lobes of insulator phases where the occupation increases with the chemical potential. This results in a ladder of occupations where each sublattice is increasingly occupied [see Fig.~\ref{sec:PD;fig:n_0o02}]. 

As the tunneling $t$ increases, the lattice shows a combination of insulator and superfluid phases. In between the insulator lobes, the lattice shows intermediate SF/MI$_{\nu}$ phases with increasing occupation, whereas it shows a complete pSF for small $\mu$. However, if the tunneling is large enough, the lattice shows a uniform 2SF phase. We note that the transition to the 2SF for $\mu=0$ in Fig.~\ref{sec:PD;fig:PD1}(a) occurs at $t/U\approx 1.0$ (beyond the scale in the figure).

Particularly interesting are the small regions with semi-polarized superfluids (spSF). These correspond to phases where both enantiomer species are superfluid, but one has a larger occupation than the other. These regions simply connect the pSF and SF/MI phases with the uniform 2SF phase. The spSF phases have energy differences per site with adjacent phases up to the 10\%, with a difference of up to $\Delta E/N\approx 0.02U$ for the lower spSF phase. However, we stress that the Gutzwiller approach is mean-field in nature, and thus the predictions for small phases should not be considered accurate.
In addition, polarized phases in extended Bose-Hubbard models suffer from competing metastable states~\cite{menotti_metastable_2007}. Therefore systems with large asymmetries between the chemical potentials might be necessary to observe an spSF phase.

For intermediate dipolar strengths $V_{LR}>U/4$ and $V/2^{2/3}<U/4$, the lattice polarizes with no uniform phases [see Fig.~\ref{sec:PD;fig:PD1}(b)]. In this case, the insulator lobes simply correspond to pMI and pDW phases with increasing occupation. For larger tunneling, the lattice shows a pSF phase, as expected. However, the pDW lobes are surrounded by small pSS phases, which are not present for weaker dipolar repulsion. A similar behavior is obtained with extended Bose-Hubbard models~\cite{iskin_route_2011}.

For strong dipolar repulsion between all molecules [see Fig.~\ref{sec:PD;fig:PD1}(c)] there is a complete polarization to one of the four sublattices for small tunneling. Therefore, all insulator lobes correspond to pDW$_\nu$ phases with increasing occupation. Moreover, these lobes are constrained to chemical potentials of $\nu-1<\mu/U<\nu$. The insulator lobes are also surrounded by a continuous pSS phase. However, for large tunneling the lattice still shows a pSF phase. This is because the repulsion between left and right enantiomers has a stronger effect than the repulsion between enantiomers of the same handedness due to the shorter distance between left and right sites.

To illustrate how the occupations change in the different phases,  in Fig.~\ref{sec:PD;fig:n_0o02} we show average occupations for fixed tunnelings for $V=V_{LR}=0.2U$. The phases can be recognized by following Tables~\ref{sec:phases;table:phasesMI} and~\ref{sec:phases;table:phasesSF}. Note that for zero tunneling the occupations show only multiples of quarters of integers, signaling insulator phases. In contrast, for finite tunneling the occupations are continuous. This signals superfluid phases between the insulator lobes. 

\begin{figure}[t]
\centering
\includegraphics[scale=0.8]{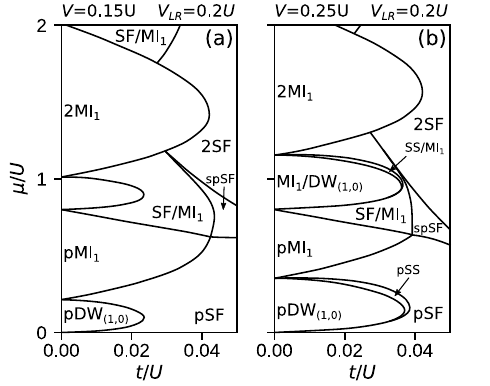}
\caption{Phase diagrams for $V\neq V_{LR}$ and $\mu_L=\mu_R$ as a function of the chemical potential $\mu/U$ and hopping $t/U$.}
\label{sec:PD;fig:PD2}
\end{figure}

Finally, in Fig.~\ref{sec:PD;fig:PD2} we show diagrams for $V\neq V_{LR}$. We show diagrams for $V_{LR}=0.2U$ to compare with Fig.~\ref{sec:PD;fig:PD1}(a). In general, a small change in $V$ roughly maintains the shape of the phase diagrams. A smaller $V$ shrinks the DW phases [Fig.~\ref{sec:PD;fig:PD2}(a)], whereas a larger $V$ enlarges them [Fig.~\ref{sec:PD;fig:PD2}(b)], as expected.  In addition, a larger $V$ produces small pSS phases surrounding the DW lobes, similar to what is observed in Fig.~\ref{sec:PD;fig:PD1}b.  However, note that noticeable differences between $V$ and $V_{LR}$ might not be realizable with chiral molecules as interaction energies between L-L/R-R pairs differ from those between L-R pairs at the percent level or less \cite{craig1998}.

\section{Phase separation}
\label{sec:PS}

We have studied the phase diagram in the grand-canonical ensemble where each point in the diagram has a fixed chemical potential. However, in an experiment we would aim to control the number of left and right enantiomers. In this case, instead of polarization, the system would show phase separation, where each enantiomer species occupy different regions of the lattice. This is similar to the phase separation shown by
strongly-repulsive bosonic mixtures~\cite{altman_phase_2003}.

The phase separation of enantiomers can be seen as a method of chiral discrimination. Indeed, by controlling the parameters in the system, a helicity lattice could separate two enantiomer species. This could provide a way to control and study enantiomers of a specific handedness in the ultracold regime.

\begin{figure}[t]
\centering
\includegraphics[scale=0.78]{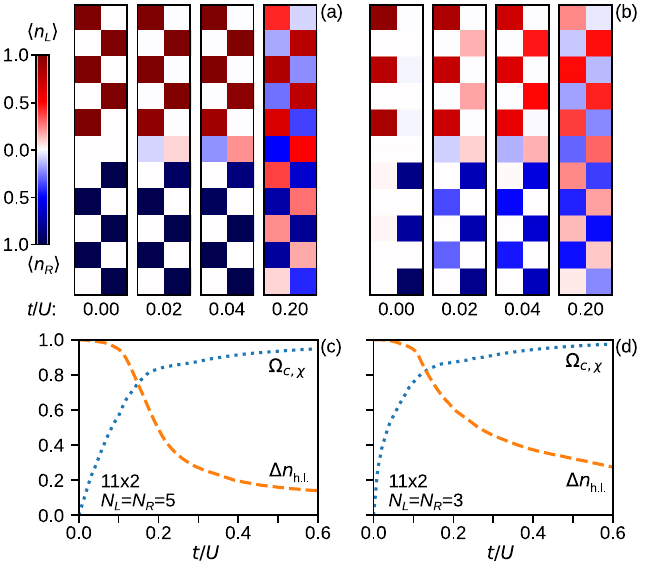}
\caption{Ground-state occupations $\langle n_\chi \rangle$ [(a-b)] and condensate fractions $\Omega_{c,\chi}$ and polarizations $\Delta n_\text{h.l.}$ [(c-d)] obtained from ED for $V=V_{LR}=0.2U$. We consider five enantiomers of each handedness in (a) and (c), and three in (b) and (d).}
\label{sec:PS;fig:ED}
\end{figure}

To illustrate such phase separation, we have performed exact diagonalizations (ED)~\cite{raventos_cold_2017} of the Hamiltonian (\ref{sec:model;eq:H}) for a small fixed number of left and right enantiomers. We show examples of different occupations in Fig.~\ref{sec:PS;fig:ED}(a-b), where we have chosen $V=V_{LR}=0.2U$ to compare with Fig.~\ref{sec:PD;fig:PD1}(a). In addition, to better illustrate the results, we have employed a 9$\times$2 lattice where the $x$-axis is periodic and the $y$-axis is finite.

For small tunneling $t/U$, Fig.~\ref{sec:PS;fig:ED}(a) shows two phase-separated MI, instead of a polarized one. As the tunneling increases, the two enantiomer species start occupying both halves of the lattice, showing a 2SF phase. Similarly, Fig.~\ref{sec:PS;fig:ED}(b) shows two phase-separated DW for small $t/U$, while for intermediate tunneling both enantiomers remain phase-separated, but become superfluid in analogy to a pSF phase. Finally, for large tunneling Fig.~\ref{sec:PS;fig:ED}(b) shows two homogeneous superfluids, as expected.

To better illustrate the phases in our ED calculations, in Fig.~\ref{sec:PS;fig:ED}(c-d) we show the condensate fraction $\Omega_{c,\chi}$ (blue lines) of each enantiomer species~\cite{zhang_exact_2010}. We observe that in both cases, the molecules show a significant condensation for $t/U\gtrsim 0.2$. Moreover, $\Omega_{c,\chi}$ increases slightly more rapidly in (d), consistent with the smaller pDW lobe in Fig.~\ref{sec:PD;fig:PD1}(a). In addition, to quantify the level of phase separation, we introduce the polarization parameter $\Delta n_\textrm{h.l.}=|N_{\chi,\text{u}}-N_{\chi,\text{b}}|/(N_{\chi,\text{u}}+N_{\chi,\text{b}})$, where $N_{\chi,\text{u}}$ and $N_{\chi,\text{b}}$ are the number of enantiomer $\chi$ in the upper and bottom halves of the lattice, respectively. The polarization (orange lines) decreases with $t$, showing that for large tunneling the system converges to a homogeneous phase. In both cases there is a significant polarization for $t/U\lesssim 0.2$, above the transitions reported in Fig.~\ref{sec:PD;fig:PD1}(a) for small $\mu$. Nevertheless, we observe that $\Delta n_\textrm{h.l.}$ decreases more rapidly in (c), consistent with the smaller polarized phases for larger $\mu$.

Our ED calculations confirm the presence of the most prominent phases shown in Fig.~\ref{sec:PD;fig:PD1}, including polarized and homogeneous MI, DW, and SF phases. We again stress that the Gutzwiller ansatz does not take quantum fluctuations into account, which can significantly change the phase diagram. Nevertheless, even though we observe signatures of the different phases with ED, calculations for small lattices are not able to unambiguously locate phase transitions~\cite{raventos_cold_2017}. Therefore, the Gutzwiller calculations should be contrasted instead with more sophisticated many-body approaches in the future.

\section{Conclusions}

This work is a proof-of-principle, describing completely new quantum phases in the recently proposed helicity lattices.
We have shown that repulsive dipolar interactions between chiral molecules immersed in helicity lattices can induce a plethora of quantum phases. A strong dipolar repulsion between molecules induces phases with left/right polarization, as well as phases with asymmetric occupations. Future experiments could produce these systems and examine these novel forms of chiral matter. 

Future work will include consideration of realistic interactions between molecules~\cite{salam_effect_2006} as well as effects from their internal structure~\cite{Carr_2009,wall2015quantum} such as from molecular rotation~\cite{wall2015quantum,di2020rotational,koch_quantum_2019,dawid_two_2018}. In this direction, Hubbard models for ultracold diatomic molecules have been proposed~\cite{Wall_2009diat,wall2015quantum,docaj_ultracold_2016, wall_microscopic_2017}. In addition, we intend to employ beyond mean-field approaches, such as Quantum Monte-Carlo~\cite{prokofev_exact_1998} or the Quantum Gutzwiller approach~\cite{caleffi_quantum_2020}, to provide a more accurate description of helicity lattices. This will also enable us to correctly study lattices with different geometries~\cite{van_kruining_superpositions_2018}.

Ultracold chiral molecules have been proposed as good candidates to test parity violation~\cite{bargueno_parity_2009,bast_analysis_2011,isaev_polyatomic_2016,cournol_new_2019,augenbraun_molecular_2020} - helicity lattices could provide a better control of cold chiral molecules. In the future, we intend to study Hubbard-like models with energy differences between enantiomers. This will enable us to narrow the conditions in which helicity lattices could probe parity violation, bringing us a step closer to an understanding of this fundamental effect.

\begin{acknowledgments}
  JBG acknowledges initial discussions with Rob Cameron.
  We acknowledge funding from EPSRC (UK) through Grant No. EP/V048449/1 and the Leverhulme Trust.
\end{acknowledgments}

\bibliography{biblio}

\end{document}